\documentclass[a4paper,twocolumn]{article}
\usepackage{graphicx}
\usepackage{indentfirst}
\usepackage{float}
\usepackage{fancyhdr}
\usepackage{amssymb}
\usepackage{amsmath}
\usepackage{cite}
\usepackage[labelfont=bf]{caption}
\usepackage{titlesec}

\usepackage[tmargin=1.9cm,lmargin=1.9cm,rmargin=2.54cm,bmargin=2.54cm,
            headheight=23.2pt,headsep=0.4cm]{geometry}
\newcommand{\HRule}{\rule{\linewidth}{0.3mm}}
\usepackage{fancyhdr}
\fancypagestyle{arci}{
  \fancyhf{}
  \fancyhead[C]{\footnotesize{\textit{1st IFSA Winter Conference on Automation,
                                      Robotics \& Communications for Industry 4.0
                                      (ARCI' 2021),\\
                                      17--19 Feb 2021 Chamonix-Mont-Blanc,
                                      France}}}
}

\usepackage[dvipsnames]{xcolor}
\usepackage[colorlinks]{hyperref}
\allowdisplaybreaks

\titleformat*{\section}{\large\bfseries}
\titleformat*{\subsection}{\large\bfseries}
\titlelabel{\thetitle.\ }

\makeatletter
\g@addto@macro\@openbib@code{\setlength{\itemsep}{0pt}}
\makeatother
\begin{document}
\twocolumn[\begin{@twocolumnfalse}
Oral \hfill{ } Topic: Communications
\begin{center}
\Large\textbf{Turbo Coded Single User Massive MIMO with Precoding}
\end{center}

\begin{center}
{\textbf{K. Vasudevan, Gyanesh Kumar Pathak, A. Phani Kumar Reddy}}\\
\small{Department of Electrical Engineering, Indian Institute of Technology Kanpur-208016, India.}\\
\small{\{vasu, pathak, phani\}@iitk.ac.in}
\end{center}

\HRule \\
\begin{footnotesize}
\textbf{Summary:} Precoding is a method of compensating the channel at the
transmitter. This work presents a novel method of data detection in turbo coded
single user massive multiple input multiple output (MIMO) systems using precoding.
We show via computer simulations that, when precoding is used, re-transmitting the
data does not result in significant reduction in bit-error-rate (BER), thus
increasing the spectral efficiency, compared to the case without precoding. Moreover,
increasing the number of transmit and receive antennas results in improved BER.

\vspace{\baselineskip}
\end{footnotesize}
\HRule
\vspace{0.5cm}

\end{@twocolumnfalse}]

\section{Introduction}
\label{Sec:Intro}
Precoding at the transmitter is a technique that dates back to the era of voiceband
modems or wired communications \cite{6768299,49851,93096,93098,120349,119688,237880}.
The term ``precoding'' is quite generic and refers to one or more of the many
different functionalities, as given below:
\begin{enumerate}
    \item It compensates for the distortion introduced by the channel. Note that
          channel compensation at the receiver is referred to as equalization
          \cite{6773303,1457566,1182531,VASUDEVAN20042271,Singer04,Vasu07,Vasu_Book10}.
          Here, channel compensation implies
          removal or minimization of intersymbol interference (ISI).
    \item It performs error control coding, besides channel compensation.
    \item It shapes the spectrum of the transmitted signal, and renders it
          suitable for propagation over the physical channel. Note that most
          channels do not propagate a dc signal and precoding is used to remove
          the dc component in the message signal. At this point, it is important
          to distinguish between a message signal and the transmitted signal.
\end{enumerate}
In the context of wireless multiple input, multiple output (MIMO) systems, the
main task of the precoder is to remove interchannel interference (ICI), either for
single-user or multi-user case
\cite{8169014,9024294,9015969,9025051,9027848,9031293,9040266}. It should
be observed that precoding requires knowledge of the channel state information (CSI)
at the transmitter, which is usually fed back by the receiver to the transmitter. The
receiver estimates CSI from a known training signal that is sent by the transmitter.
CSI usually refers to the channel impulse response (CIR) or its statistics
(mean and covariance), depending on the type of precoder used. Thus,
precoding requires the channel to be time invariant or wide sense
stationary (WSS) over at least one transmit and receive duration. Moreover, precoding
can only be performed on systems employing time division duplex (TDD),
which is a method of half duplex telecommunication. In other words, the channel needs
to be reciprocal, that is, the CIR from the transmitter to receiver must be
identical to that from receiver to transmitter.

In this work, we describe an elegant precoding method which reduces ICI in single
user massive MIMO systems and compare it with the case without precoding
\cite{KV_OpSigPJ2019,Vasu_intech:2019}. Rayleigh flat fading channel is assumed.
If the channel is frequency selective, orthogonal frequency division multiplexing (OFDM)
can be used
\cite{Vasu_Book10,6514242,6663392,6663417,6685594,Vasudevan2015,Vasu_ICWMC2016,
Vasu_Adv_Tele_2017,Vasu_intech:2019,c7888430-cbc2-4c14-87bb-b52780478d85,
73ddc0ea-7d42-4fdd-969d-da08c8e4d0c0,d4bbbdf0-7468-4727-9ebe-76d5e6160b64,
KV_SSID2020,9286489e-9169-41d7-9950-1f89bb42fd15}.

This work is organized as follows. Section~\ref{Sec:Signal_Model} describes the
signal model. In Section~\ref{Sec:Precoding} precoding for single user massive
MIMO is discussed. Section~\ref{Sec:Results} presents the simulation results
and conclude the work in Section~\ref{Sec:Conclude}.
\section{Signal Model}
\label{Sec:Signal_Model}
\begin{figure*}[tbhp]
\begin{center}
\input{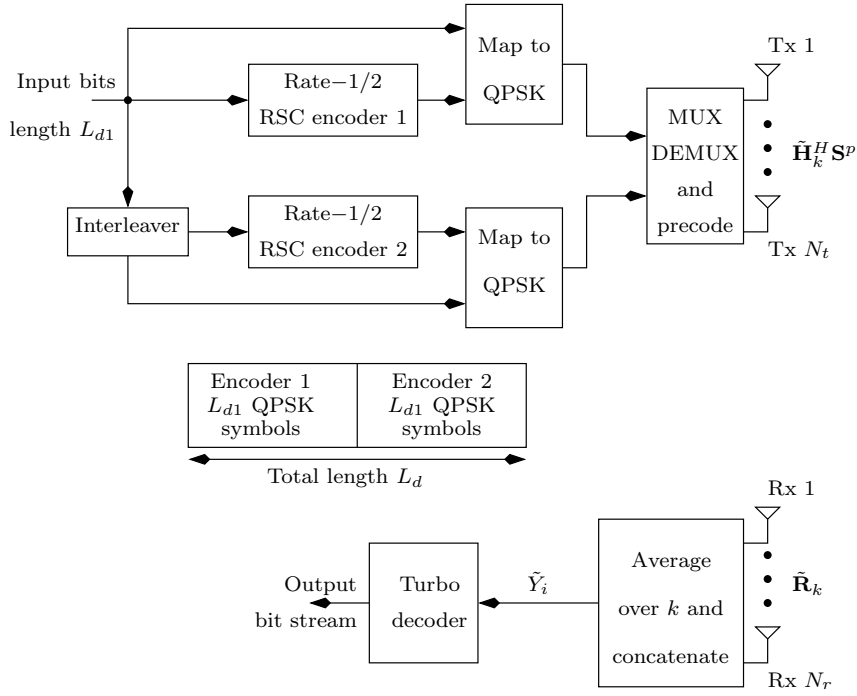}
\caption{System model.}
\label{Fig:Pap14_System}
\end{center}
\end{figure*}

Consider a precoded MIMO system with $N_t$ transmit and $N_r$ receive antennas, as
shown in \textbf{Fig.}~\ref{Fig:Pap14_System} \cite{KV_OpSigPJ2019}.
The precoded received signal in the $k^{th}$ ($0\leq k\leq N_{rt}-1$,
$k$ is an integer), re-transmission is given by
\begin{equation}
\label{Eq:Pre_Massive_MIMO_Eq1}
\tilde{\mathbf{R}}_k = \tilde{\mathbf{H}}_k
                       \tilde{\mathbf{H}}_k^H
                       \mathbf{S}^p +
                       \tilde{\mathbf{W}}_k
\end{equation}
where $\tilde{\mathbf{R}}_k\in \mathbb{C}^{N_r\times 1}$ is the received
vector, $\tilde{\mathbf{H}}_k\in \mathbb{C}^{N_r\times N_t}$ is the channel
matrix and $\tilde{\mathbf{W}}_k\in \mathbb{C}^{N_r\times 1}$ is the additive 
white Gaussian noise (AWGN) vector. The transmitted symbol vector is
$\mathbf{S}^p\in \mathbb{C}^{N_r\times 1}$, whose elements are drawn from an
$M$-ary constellation. Boldface letters denote vectors or matrices. Complex
quantities are denoted by a  tilde. However tilde is not used for complex
symbols $\mathbf{S}^p$. The elements of $\tilde{\mathbf{H}}_{k}$ are
statistically independent, zero mean, circularly symmetric complex Gaussian with
variance per dimension equal to
$\sigma_{H}^{2}$, as given by (2) of \cite{KV_OpSigPJ2019}. Similarly, the elements of
$\tilde{\mathbf{W}}_k$ are statistically independent, zero mean, circularly
symmetric complex Gaussian with 
variance per dimension equal to $\sigma_W^2$, as given by (3) of \cite{KV_OpSigPJ2019}.

In this work, the elements of $\mathbf{S}^p$ are turbo coded and mapped to a
QPSK constellation with coordinates $\pm 1 \pm\mathrm{j}$, as depicted
in \textbf{Fig.}~\ref{Fig:Pap14_System}.
Moreover, here $\tilde{\mathbf{H}}_k$ is an
$N_r\times N_t$ matrix, whereas in \cite{KV_OpSigPJ2019} $\tilde{\mathbf{H}}_k$ is
an $N\times N$ matrix. We assume that $\tilde{\mathbf{H}}_k$ and
$\tilde{\mathbf{W}}_k$ are independent across re-transmissions, hence (4) in 
\cite{KV_OpSigPJ2019} is valid with $N$ replaced by $N_r$.
We now proceed to analyze the signal model in (\ref{Eq:Pre_Massive_MIMO_Eq1}).
\section{Precoding}
\label{Sec:Precoding}
The $i^{th}$ element of $\tilde{\mathbf{R}}_k$ in (\ref{Eq:Pre_Massive_MIMO_Eq1}) is
\begin{equation}
\label{Eq:Pre_Massive_MIMO_Eq3}
\tilde{R}_{k,\, i} = \tilde{F}_{k,\, i,\, i} S_i +
                     \tilde{I}_{k,\, i} +
                     \tilde{W}_{k,\, i}
                     \quad \mbox{for $1\leq i\leq N_r$}
\end{equation}
where
\begin{align}
\label{Eq:Pre_Massive_MIMO_Eq4}
\tilde{F}_{k,\, i,\, i} & = \sum_{j=1}^{N_t}
                            \left|
                            \tilde{H}_{k,\, i,\, j}
                            \right|^2                         \nonumber  \\ 
\tilde{I}_{k,\, i}      & = \sum_{\substack{j=1\\j\neq i}}^{N_r}
                            \tilde{F}_{k,\, i,\, j} S_j       \nonumber  \\
\tilde{F}_{k,\, i,\, j} & = \sum_{l=1}^{N_t}
                            \tilde{H}_{k,\, i,\, l}
                            \tilde{H}_{k,\, j,\, l}^* \qquad \mbox{for $i\ne j$}.
\end{align}
The desired signal in (\ref{Eq:Pre_Massive_MIMO_Eq3}) is $F_{k,\, i,\, i}S_i$, the
interference term is $\tilde{I}_{k,\, i}$ and the noise term is
$\tilde{W}_{k,\, i}$. Now
\begin{align}
\label{Eq:Pre_Massive_MIMO_Eq5}
 E
\left[
\tilde{F}_{k,\, i,\, i}^2
\right] & =  E
            \left[
            \sum_{j=1}^{N_t}
            \left|
            \tilde{H}_{k,\, i,\, j}
            \right|^2
            \sum_{l=1}^{N_t}
            \left|
            \tilde{H}_{k,\, i,\, l}
            \right|^2
            \right]                              \nonumber  \\
        & =  E
            \left[
            \sum_{j=1}^{N_t}
            \tilde{H}_{k,\, i,\, j,\, I}^2 +
            \tilde{H}_{k,\, i,\, j,\, Q}^2
            \right.                              \nonumber  \\
        &   \qquad
            \times \left.
            \sum_{l=1}^{N_t}
            \tilde{H}_{k,\, i,\, l,\, I}^2 +
            \tilde{H}_{k,\, i,\, l,\, Q}^2
            \right]                              \nonumber  \\
        & = 4\sigma_H^4 N_t (N_t+1)
\end{align}
where the subscript ``$I$'' denotes the in-phase part and the subscript ``$Q$''
denotes the quadrature part of a complex quantity and the following relation
has been used \cite{Papoulis91,Vasu_AC_PS}
\begin{equation}
\label{Eq:Pre_Massive_MIMO_Eq6}
 E
\left[
 X^4
\right] = 3 \sigma_X^4
\end{equation}
where $X$ is a zero-mean, real-valued Gaussian random variable with variance
$\sigma^2_X$. Moreover from (\ref{Eq:Pre_Massive_MIMO_Eq4}) and (2) in
\cite{KV_OpSigPJ2019}
\begin{equation}
\label{Eq:Pre_Massive_MIMO_Eq6_1}
 E
\left[
\tilde{F}_{k,\, i,\, i}
\right] = 2 \sigma_H^2 N_t.
\end{equation}
\begin{figure*}[tbhp]
\begin{center}
\input{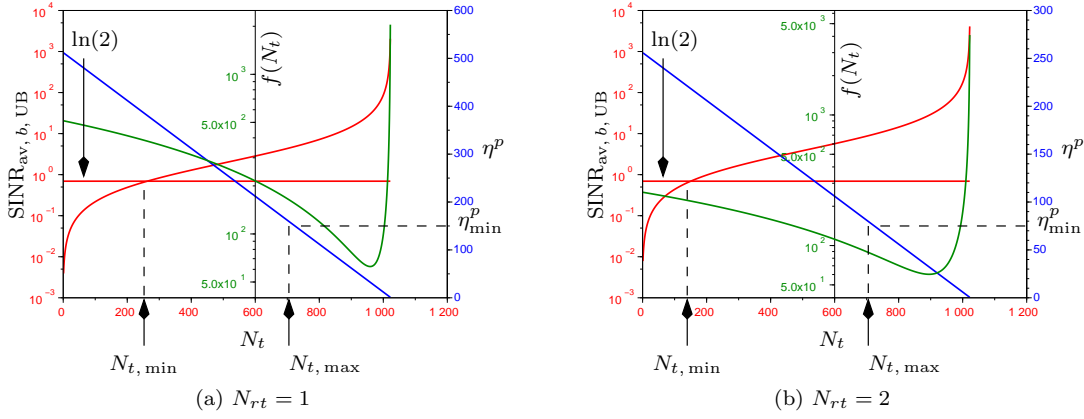}
\caption{$\mathrm{SINR}_{\mathrm{av},\, b,\,\mathrm{UB}}$ and $\eta^p$ as a
         function of $N_t$ for $N_{\mathrm{tot}}=1024$.}
\label{Fig:Pre_SINR_UB_Eta_p}
\end{center}
\end{figure*}
We also have
\begin{align}
\label{Eq:Pre_Massive_MIMO_Eq7}
 E
\left[
\left|
 \tilde{I}_{k,\, i}
\right|^2
\right] & =  E
            \left[
            \sum_{\substack{j=1\\j\neq i}}^{N_r}
            \tilde{F}_{k,\, i,\, j} S_j
            \right.                                     \nonumber  \\
        &   \qquad \times
            \left.
            \sum_{\substack{l=1\\l\neq i}}^{N_r}
            \tilde{F}_{k,\, i,\, l}^* S_l^*
            \right]                                     \nonumber  \\
        & = \sum_{\substack{j=1\\j\ne i}}^{N_r}
            \sum_{\substack{l=1\\l\ne i}}^{N_r}
             P_{\mathrm{av}}
             E
            \left[
            \tilde{F}_{k,\, i,\, j}
            \tilde{F}_{k,\, i,\, l}^*
            \right]
            \delta_K(j-l)                               \nonumber  \\
        & =  P_{\mathrm{av}}
            \sum_{\substack{j=1\\j\ne i}}^{N_r}
             E
            \left[
            \left|
            \tilde{F}_{k,\, i,\, j}
            \right|^2
            \right]
\end{align}
where $\delta_K(\cdot)$ is the Kronecker delta function 
\cite{Vasu_Book10,KV_OpSigPJ2019}, we have assumed
independence between $\tilde{F}_{k,\, i,\, j}$ and $S_j$ and \cite{KV_OpSigPJ2019}
\begin{align}
\label{Eq:Pre_Massive_MIMO_Eq8}
 E
\left[
 S_j S_l^*
\right] & = P_{\mathrm{av}} \delta_K(j-l)    \nonumber  \\
        & = 2 \delta_K(j-l).
\end{align}
Now
\begin{align}
\label{Eq:Pre_Massive_MIMO_Eq9}
 E
\left[
\left|
\tilde{F}_{k,\, i,\, j}
\right|^2
\right] & =  E
            \left[
            \sum_{l=1}^{N_t}
            \tilde{H}_{k,\, i,\, l}
            \tilde{H}_{k,\, j,\, l}^* 
            \right.                              \nonumber  \\
        &   \qquad \times
            \left.
            \sum_{m=1}^{N_t}
            \tilde{H}_{k,\, i,\, m}^*
            \tilde{H}_{k,\, j,\, m}
            \right]                              \nonumber  \\
        & = \sum_{l=1}^{N_t}
            \sum_{m=1}^{N_t}
             E
            \left[
            \tilde{H}_{k,\, i,\, l}
            \tilde{H}_{k,\, i,\, m}^*
            \right]                              \nonumber  \\
        &   \qquad \times
             E
            \left[
            \tilde{H}_{k,\, j,\, m}
            \tilde{H}_{k,\, j,\, l}^*
            \right]                              \nonumber  \\
        & = \sum_{l=1}^{N_t}
            \sum_{m=1}^{N_t}
             4
            \sigma_H^4
            \delta_K(l-m)                        \nonumber  \\
        & =  4
            \sigma_H^4 N_t.
\end{align}
Substituting (\ref{Eq:Pre_Massive_MIMO_Eq9}) in (\ref{Eq:Pre_Massive_MIMO_Eq7})
and using (\ref{Eq:Pre_Massive_MIMO_Eq8}) we get
\begin{equation}
\label{Eq:Pre_Massive_MIMO_Eq10}
 E
\left[
\left|
\tilde{I}_{k,\, i}
\right|^2
\right] = 8 \sigma_H^4 N_t (N_r-1).
\end{equation}
Due to independence between $\tilde{I}_{k,\, i}$ and $\tilde{W}_{k,\, i}$ in
(\ref{Eq:Pre_Massive_MIMO_Eq3}) we have from (\ref{Eq:Pre_Massive_MIMO_Eq10})
and (3) of \cite{KV_OpSigPJ2019}
\begin{align}
\label{Eq:Pre_Massive_MIMO_Eq11}
 E
\left[
\left|
\tilde{I}_{k,\, i} + \tilde{W}_{k,\, i}
\right|^2
\right] & =  E
            \left[
            \left|
            \tilde{I}_{k,\, i}
            \right|^2
            \right] +
            E
            \left[
            \left|
            \tilde{W}_{k,\, i}
            \right|^2
            \right]                                    \nonumber  \\
        & = 8 \sigma_H^4 N_t (N_r-1) + 2\sigma_W^2     \nonumber  \\
        & = \sigma^2_{U'}                              \qquad \mbox{(say)}.
\end{align}
Now, each element of $\mathbf{S}^p$ in (\ref{Eq:Pre_Massive_MIMO_Eq1}) carries
$1/(2N_{rt})$ bits of information \cite{KV_OpSigPJ2019}. Therefore, each element
of $\tilde{\mathbf{R}}_k$ also carries $1/(2N_{rt})$ bits of information. Hence,
the average signal to interference plus noise ratio per bit of $\tilde{R}_{k,\, i}$ in
(\ref{Eq:Pre_Massive_MIMO_Eq3}) is defined as, using (\ref{Eq:Pre_Massive_MIMO_Eq5}),
(\ref{Eq:Pre_Massive_MIMO_Eq8}) and (\ref{Eq:Pre_Massive_MIMO_Eq11})
\begin{align}
\label{Eq:Pre_Massive_MIMO_Eq12}
\mathrm{SINR}_{\mathrm{av},\, b}
        & = \frac{
             E
            \left[
            \left|
            \tilde{F}_{k,\, i,\, i} S_i
            \right|^2
            \right]
            \times 2N_{rt}}
            {
             E
           \left[
           \left|
           \tilde{I}_{k,\, i} + \tilde{W}_{k,\, i}
           \right|^2
           \right]
           }                                               \nonumber  \\
        & = \frac{8\sigma_H^4 N_t (N_t+1)\times 2 N_{rt}}
            {
             8\sigma_H^4 N_t (N_r-1) + 2\sigma_W^2
            }.
\end{align}
\begin{figure*}[tbhp]
\begin{center}
\input{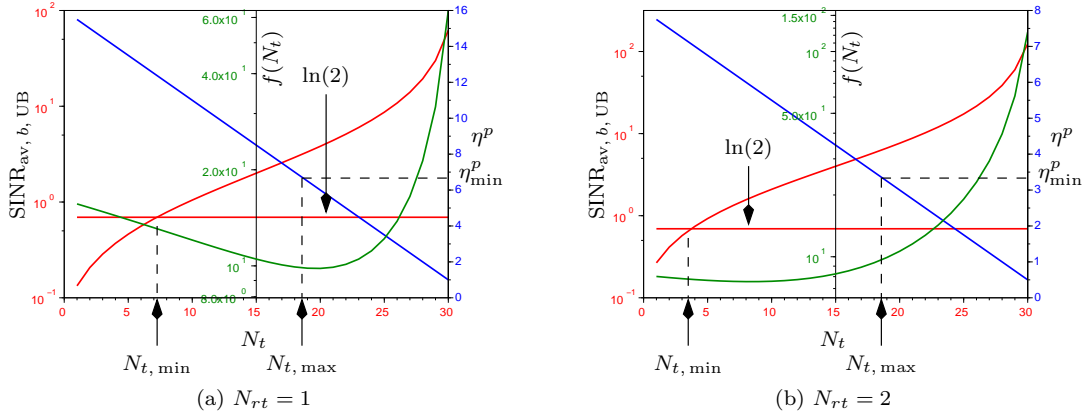}
\caption{$\mathrm{SINR}_{\mathrm{av},\, b,\,\mathrm{UB}}$ and $\eta^p$ as a
         function of $N_t$ for $N_{\mathrm{tot}}=32$.}
\label{Fig:Pre_SINR_UB_Eta_p_Ntot32}
\end{center}
\end{figure*}
When $\sigma_W^2=0$ in (\ref{Eq:Pre_Massive_MIMO_Eq12}), we get the upper bound
on $\mathrm{SINR}_{\mathrm{av},\, b}$ as given below
\begin{align}
\label{Eq:Pre_Massive_MIMO_Eq13}
\mathrm{SINR}_{\mathrm{av},\, b,\,\mathrm{UB}}
& = \frac{8\sigma_H^4 N_t (N_t+1)\times 2 N_{rt}}
         {8\sigma_H^4 N_t (N_r-1)}                  \nonumber  \\
& = \frac{2N_{rt}(N_t+1)}{N_r-1}.
\end{align}
The information contained in $\mathbf{S}^p$ in (\ref{Eq:Pre_Massive_MIMO_Eq1})
is $N_r/(2N_{rt})$ bits. Hence the spectral efficiency of the precoded system is
\begin{equation}
\label{Eq:Pre_Massive_MIMO_Eq14}
\eta^p = \frac{N_r}{2N_{rt}} \qquad \mbox{bits per transmission}.
\end{equation}
Note that both (\ref{Eq:Pre_Massive_MIMO_Eq13}) and (\ref{Eq:Pre_Massive_MIMO_Eq14})
need to be as large as possible to minimize the BER and maximize the spectral
efficiency. Let
\begin{equation}
\label{Eq:Pre_Massive_MIMO_Eq15}
N_{\mathrm{tot}} = N_t+N_r.
\end{equation}
Define
\begin{align}
\label{Eq:Pre_Massive_MIMO_Eq16}
f(N_t) & = \mathrm{SINR}_{\mathrm{av},\, b,\,\mathrm{UB}} + \eta_p  \nonumber  \\
       & = \frac{2N_{rt}(N_t+1)}{N_r-1} + \frac{N_r}{2N_{rt}}       \nonumber  \\
       & = \frac{2N_{rt}(N_t+1)}{N_{\mathrm{tot}}-N_t-1} +
           \frac{N_{\mathrm{tot}}-N_t}{2N_{rt}}
\end{align}
where we have used (\ref{Eq:Pre_Massive_MIMO_Eq15}).
We need to find $N_t$ such that $f(N_t)$ is maximized. The plot of
$\mathrm{SINR}_{\mathrm{av},\, b,\,\mathrm{UB}}$ (red curve), $\eta^p$ (blue curve)
and $f(N_t)$ (green curve), as a function of $N_t$, keeping $N_{\mathrm{tot}}$
fixed, is depicted in \textbf{Fig.}~\ref{Fig:Pre_SINR_UB_Eta_p} and
\ref{Fig:Pre_SINR_UB_Eta_p_Ntot32}. Note that
$\mathrm{SINR}_{\mathrm{av},\, b,\,\mathrm{UB}}$ increases monotonically and $\eta^p$
decreases monotonically, with increasing $N_t$. We also find that
$f(N_t)$ has a minimum (not maximum) at
\begin{equation}
\label{Eq:Pre_Massive_MIMO_Eq17}
N_t = N_{\mathrm{tot}}-2N_{rt}\sqrt{N_{\mathrm{tot}}} -1
\end{equation}
which is obtained by differentiating $f(N_t)$ in (\ref{Eq:Pre_Massive_MIMO_Eq16})
with respect to $N_t$ and setting the result to zero. Therefore, the only possible
solution is to avoid the minimum.  Clearly we require
$\mathrm{SINR}_{\mathrm{av},\, b,\,\mathrm{UB}}>\ln(2)$, since it is the minimum
average SNR per bit required for error-free transmission over any type of channel
\cite{KV_OpSigPJ2019}. We also require $\eta^p > \eta^p_{\mathrm{min}}$, where
$\eta^p_{\mathrm{min}}$ is chosen by the system designer. Thus, we arrive at a
range of the number of transmit antennas
($N_{t,\,\mathrm{min}} \le N_t \le N_{t,\,\mathrm{max}}$) that can be used, as shown
in \textbf{Fig.}~\ref{Fig:Pre_SINR_UB_Eta_p} and
\ref{Fig:Pre_SINR_UB_Eta_p_Ntot32}. Note that in
\textbf{Fig.}~\ref{Fig:Pre_SINR_UB_Eta_p_Ntot32}(b) the minimum of
$f(N_t)$ cannot be avoided, since $\eta^p_{\mathrm{min}}$ would be too small.

Next, similar to (20) in \cite{KV_OpSigPJ2019}, consider
\begin{align}
\label{Eq:Pre_Massive_MIMO_Eq18}
\tilde{Y}_i & = \frac{1}{N_{rt}}
                \sum_{k=0}^{N_{rt}-1}
                \tilde{R}_{k,\, i}                      \nonumber  \\
            & = \frac{1}{N_{rt}}
                \sum_{k=0}^{N_{rt}-1}
                \left(
                \tilde{F}_{k,\, i,\, i} S_i +
                \tilde{I}_{k,\, i} +
                \tilde W_{k,\, i}
                \right)                                \nonumber  \\
            & =  F_i S_i + \tilde{U}_i \qquad \mbox{for $1\le i \le N_r$}
\end{align}
where $\tilde{R}_{k,\, i}$ is given by (\ref{Eq:Pre_Massive_MIMO_Eq3}), $F_i$
is real-valued and
\begin{align}
\label{Eq:Pre_Massive_MIMO_Eq19}
F_i         & = \frac{1}{N_{rt}}
                \sum_{k=0}^{N_{rt}-1}
                \tilde{F}_{k,\, i,\, i}                        \nonumber  \\
\tilde{U}_i & = \frac{1}{N_{rt}}
                \sum_{k=0}^{N_{rt}-1}
                \left(
                \tilde{I}_{k,\, i} +
                \tilde{W}_{k,\, i}
                \right)                                        \nonumber  \\
            & = \frac{1}{N_{rt}}
                \sum_{k=0}^{N_{rt}-1}
                \tilde{U}_{k,\, i}'   \qquad \mbox{(say)}.
\end{align}
Since $\tilde{F}_{k,\, i,\, i}$ and $\tilde{U}_{k,\, i}'$ are statistically
independent over re-transmissions ($k$), we have
\begin{align}
\label{Eq:Pre_Massive_MIMO_Eq20}
 E
\left[
F_i^2
\right]     & = \frac{1}{N_{rt}^2}
                 E
                \left[
                \sum_{k=0}^{N_{rt}-1}
                \tilde{F}_{k,\, i,\, i}
                \sum_{n=0}^{N_{rt}-1}
                \tilde{F}_{n,\, i,\, i}
                \right]                                        \nonumber  \\
            & = \frac{4\sigma_H^4 N_t
                \left[N_t+1+N_t(N_{rt}-1)
                \right]}{N_{rt}}                               \nonumber  \\
            & = \frac{4\sigma_H^4 N_t(N_t N_{rt}+1)}
                     {N_{rt}}                                  \nonumber  \\
 E
\left[
\left|
\tilde{U}_i
\right|^2
\right]     & = \frac{\sigma^2_{U'}}{N_{rt}}                   \nonumber  \\
            & = \frac{8 \sigma_H^4 N_t (N_r-1) + 2\sigma_W^2}{N_{rt}}
\end{align}
where we have used (\ref{Eq:Pre_Massive_MIMO_Eq5}), (\ref{Eq:Pre_Massive_MIMO_Eq6_1}),
(\ref{Eq:Pre_Massive_MIMO_Eq11}) and the fact that
\begin{equation}
\label{Eq:Pre_Massive_MIMO_Eq21}
 E
\left[
\tilde{U}_{k,\, i}'
\right] = 0
\end{equation}
where $\tilde{U}_{k,\, i}'$ is defined in (\ref{Eq:Pre_Massive_MIMO_Eq19}). Next,
we compute the average SINR per bit for $\tilde{Y}_i$ in
(\ref{Eq:Pre_Massive_MIMO_Eq18}). Note that since $\tilde{Y}_i$ is a ``combination''
of $N_{rt}$ re-transmissions, its information content is $N_{rt}/(2N_{rt})=1/2$ bit
(recall that the information content of $\tilde{R}_{k,\, i}$ in
(\ref{Eq:Pre_Massive_MIMO_Eq18}) is $1/(2N_{rt})$ bits). Therefore
\begin{align}
\label{Eq:Pre_Massive_MIMO_Eq22}
\mathrm{SINR}_{\mathrm{av},\, b,\, C}
& = \frac{E\left[\left|F_i S_i\right|^2\right]\times 2}
         {E\left[\left|\tilde{U}_i\right|^2\right]}             \nonumber  \\
& = \frac{8\sigma_H^4 N_t (N_t N_{rt}+1)\times 2}
         {8\sigma_H^4 N_t (N_r-1)+2\sigma^2_W}
\end{align}
where the subscript ``$C$'' denotes ``after combining'' and we have used
(\ref{Eq:Pre_Massive_MIMO_Eq8}) and (\ref{Eq:Pre_Massive_MIMO_Eq20}). Note that
we prefer to use the word ``combining'' rather than averaging, since it is
more appropriate in terms of the ``information content'' in $\tilde{Y}_i$. Once
again with $\sigma^2_W=0$ and $N_t N_{rt}\gg 1$ we get the approximate upper
bound on $\mathrm{SINR}_{\mathrm{av},\, b,\, C}$ as
\begin{align}
\label{Eq:Pre_Massive_MIMO_Eq23}
\mathrm{SINR}_{\mathrm{av},\, b,\, C,\,\mathrm{UB}}
& = \frac{8\sigma_H^4 N_t (N_t N_{rt}+1)\times 2}
         {8\sigma_H^4 N_t (N_r-1)}                         \nonumber  \\
& \approx
    \frac{2N_{rt}N_t}{N_r-1}                               \nonumber  \\
& \approx
    \mathrm{SINR}_{\mathrm{av},\, b,\,\mathrm{UB}}
\end{align}
when $N_t\gg 1$. Thus, the upper bound on the average SINR per bit before and
after combining are nearly identical. Observe that re-transmitting the data increases
the upper bound on the average SINR per bit, it does not improve the BER performance,
which is seen in the next section. After concatenation, the signal $\tilde Y_i$
in (\ref{Eq:Pre_Massive_MIMO_Eq18}) for $0\le i\le L_d-1$ is sent to the turbo
decoder. The details of turbo decoding will not be discussed here.
\section{Simulation Results}
\label{Sec:Results}
In this section, we discuss the results from computer simulations.
\begin{figure*}[tbhp]
\begin{center}
\input{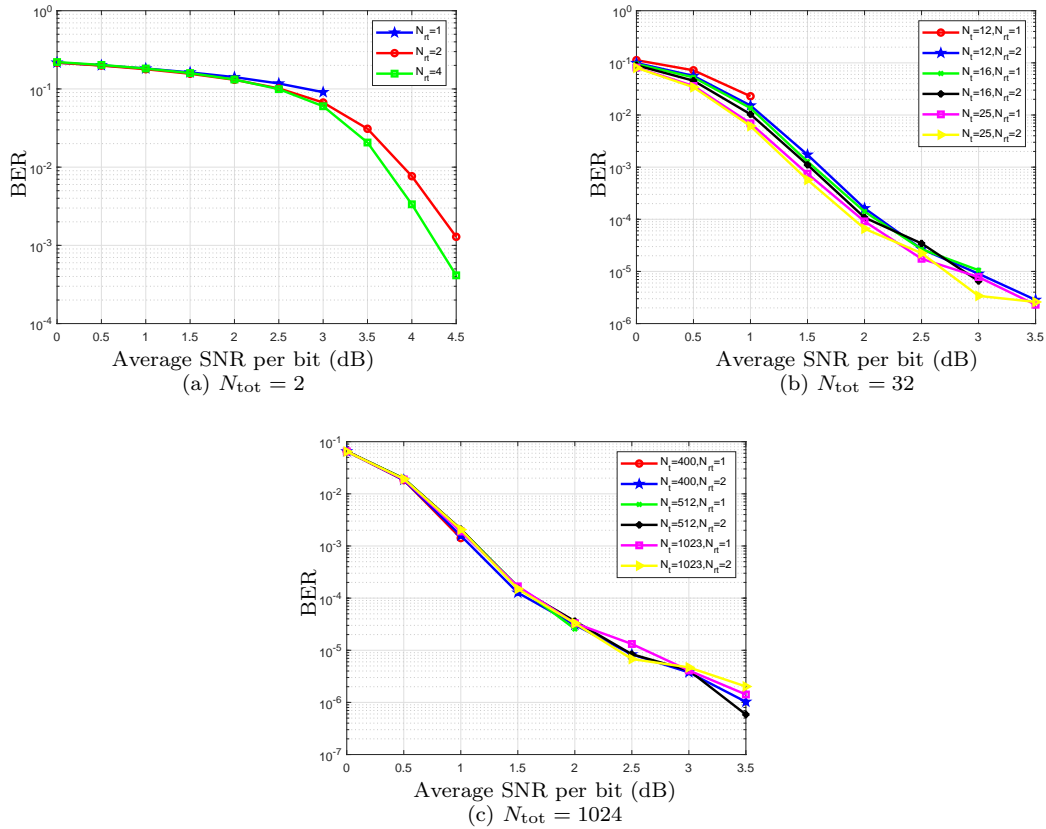}
\caption{Simulation results.}
\label{Fig:Pap14_Results}
\end{center}
\end{figure*}
The length of the data bits per ``frame'' ($L_{d1}$)  is taken to be the
smallest integer greater than 1000, which is an integer multiple of $N_r$.
Note that (see \textbf{Fig.}~\ref{Fig:Pap14_System})
\begin{equation}
\label{Eq:Pre_Massive_MIMO_Eq24}
L_d = 2 L_{d1}.
\end{equation}
The simulations were carried out over $10^4$ frames. The turbo encoder is
given by (38) of \cite{KV_OpSigPJ2019}.
\begin{itemize}
 \item \textbf{Fig.}~\ref{Fig:Pap14_Results}(a) gives the bit-error-rate (BER)
       results for a $1\times 1$ single input single output (SISO) system
       ($N_{\mathrm{tot}}=2$). We get a BER of $2\times 10^{-2}$ at an average
       SNR per bit of 3.5 dB, with $N_{rt}=4$. The corresponding spectral
       efficiency is $\eta^p=1/8$ bits per transmission. Th BER also does not
       vary significantly with the number of re-transmissions ($N_{rt}$).
 \item \textbf{Fig.}~\ref{Fig:Pap14_Results}(b) gives the results for
       $N_{\mathrm{tot}}=32$ and different combinations of transmit ($N_t$)
       and receive ($N_r$) antennas. We find that the BER is quite insensitive
       to variations in $N_t$, $N_r$ and $N_{rt}$. Moreover, the BER at an
       SNR per bit of 3.5 dB is about $2\times 10^{-6}$, which is a significant
       improvement over the SISO system. Of all the curves, $N_t=25$,
       $N_{rt}=2$ gives the lowest spectral efficiency of $\eta^p=1.75$
       bits/sec/Hz and highest
       $\mathrm{SNR}_{\mathrm{av},\, b,\,\mathrm{UB}}=12.39$ dB. Of all the
       curves, $N_t=12$, $N_{rt}=1$ gives the highest spectral efficiency
       $\eta^p=10$ bits/sec/Hz and lowest
       $\mathrm{SNR}_{\mathrm{av},\, b,\, \mathrm{UB}}=1.36$ dB.
 \item \textbf{Fig.}~\ref{Fig:Pap14_Results}(c) gives the results for
       $N_{\mathrm{tot}}=1024$ for various combinations of $N_t$, $N_r$
       and $N_{rt}$. The BER is similar to that of $N_{\mathrm{tot}}=32$. Of all
       the curves, $N_t=400$, $N_{rt}=1$ gives the highest spectral efficiency
       of $\eta^p=312$ bits/sec/Hz and lowest
       $\mathrm{SNR}_{\mathrm{av},\, b,\,\mathrm{UB}}=1.09$ dB. Of all the
       curves, $N_t=1023$, $N_{rt}=2$ gives the lowest spectral efficiency of
       $\eta^p=0.25$ and highest
       $\mathrm{SNR}_{\mathrm{av},\, b,\,\mathrm{UB}}\rightarrow\infty$.
\end{itemize}
\section{Conclusions}
\label{Sec:Conclude}
This work presents a method for data detection in turbo-coded and precoded
massive MIMO. An ideal receiver is assumed. Future work could be to simulate
a realistic precoded system with carrier and timing synchronization and
channel estimation.

\begin{footnotesize}

\bibliographystyle{IEEEtran}
\bibliography{mybib,mybib1,mybib2,mybib3,mybib4}
\end{footnotesize}
\end{document}